\documentclass{ws-ijmpcs}

\begin{document}

\markboth{C. Sigismondi}
{Solar astrometry: the status of art in 2011}

%
\catchline{}{}{}{}{}
%
\title{SOLAR ASTROMETRY: THE STATUS OF ART IN 2011}

\author{COSTANTINO SIGISMONDI}

\address{Sapienza University of Rome, Physics Dept. P.le Aldo Moro 5 \\
Roma, 00185, Italy\\
University of Nice-Sophia Antipolis - Dept. Fizeau (France); \\
IRSOL, Istituto Ricerche Solari di Locarno (Switzerland)\\
sigismondi@icra.it}

\maketitle

\begin{history}
\received{6 Feb 2012}
\revised{Day Month Year}
\end{history}

\begin{abstract}

Solar astrometry deals with the accurate measumerent of the solar diameter, and in general with the measurement of the shape of the Sun.
During the last decades several techniques have been developed to monitor the radius and the irradiance of the Sun: meridian transits, telescopes in drift-scan mode, solar astrolabes, balloons, and satellites dedicated to the measurements of the solar diameter, and space measurements of the total solar irradiance are now performed to know the relationship radius-luminosity for the Sun in this evolutionary stage of its life. The feedback of solar astrometry in climate studies is of paramount importance.
The status of art in the various fields of research here adressed is outlined.

\end{abstract}

    
\ccode{PACS numbers: 96.60.Q-,96.60.-j, 01.65.+g,}
    
\section{Introduction}    
The Sun experienced in the last four years the deepest minimum of activity since the Dalton minimum (1810-14). This phenomenon can affect the whole climate on the Earth in the future years. The relationship radius-luminosity will allow to know precisely the solar energetic input to our planet in the past, when the radius from ancient eclipses is available. With upgraded physical models of the Sun this relationship will be used to predict Sun-Earth connections. Solar astrometry deals with accurate measurements of the solar diameter, and it is an important indicator of solar activity.
\footnote{The parallel session of the third meeting "Galileo - Xu Guangxi" between Italian and Chinese scientists dedicated on solar physics, was held on October 11, 2011 in the Beijing National Astronomical Observatory headquartier. This event has anticipated of one month the fourth meeting between China and France on solar physics entitled "Understanding Solar Activity: advances and Changes".{\rm http://fcspw4.oca.eu/workshop\_2011/main\_1st.php}
I was called to chair this session together with prof. Jingxiu Wang in the month of March 2011.
The title proposed for this parallel session was "Solar astrometry and grand minima of activity", and here I outline the contributions of my research partners in Europe and Brazil.
The meteor shower of Draconids on October 8, 2011, was also observed onboard Alitalia Boeing 767 Rome-Beijing and reported in a paper.\cite{Draco}
On the Chinese side the organization was extremely efficient, and I could have a general idea of the solar physics advancements in China, sharing the whole week with the staff, kindly hosted by Prof. Wang in his office.
A visit to Huariou Solar Observing Station {\rm http://sun.bao.ac.cn/}
guided by Dr. Xiaofan Wang completed the week, and opened space for new fruitful collaborations.}

\section{Solar Diameter, Limb Darkening Function and Eclipses}

The new method to measure the solar diameter with eclipses is based on the recovering of the Limb Darkening Function's inflexion point from the Baily's beads light curve. It might solve the ambiguity of some eclipse observations made with different instruments, where the measured solar diameter was clearly dependent on the aperture of the telescope and on the density of the filter used.
Now the inflexion point is independent on the filter and the detector. The combination detector/filter should allow to measure the inflexion point and both the outer limb and the inner solar disk without saturating. The 8-bit detectors usually saturate. 12-bit or 16-bit are required. 

\section{Clavius Project on Ground-based Solar Diameter Measurements}

The debate on the metrological standards attained by ground-based solar diameter measurements is still open.
In 2008 we launched the Clavius project\cite{Arnaud} in order to investigate the possibility of a unified vision of the problem. Christopher Clavius (1535-1612) witnessed the eclipse of 1567 in Rome, object of study in the solar diameter secular changes.\cite{Clavius,Eddy,morrison}
Up to now solar astrolabes, heliometers and eclipses have been treated in different ways with respect to the measurement of the solar diameter. 

The IRSOL 45 cm Gregory-Coud\'e telescope and its twin in Iza\~{n}a,Tenerife were used by Wittmann and Bianda\cite{Wittmann} in 1990-2000 observational campaigns on solar diameter measurements with drift-scan. It has been the largest telescope used with this purpose, and new observations have been done from 2008 to 2011 in connection with solar eclipses.
S. Koutchmy and C. Bazin joined this project in 2010 by using the Carte du Ciel 33 cm refractor of the Paris Observatory.

The basic aim of the Clavius project, before monitoring the solar diameter from ground at 0.01 arcsec level, is to understand why two following measurements with drift-scan can differ each other by more than all predicted errors.

Moreover annual averages on such measurements, published since 1955 for one century of meridian transits in Greenwich and 70 years in Rome,\cite{gething}
show large scatters, up to one arcsec, not explicable with the hypothesis of Gaussian distribution of the single observations.

\subsection{Non-Gaussian seeing during solar observations}
The first results of Clavius project is to understand the role of the seeing at 0.01 Hz level.\cite{seeing} The slow motion of the solar image over the two minutes of the transit is changing continuously the value of the measured diameter. The average over several measurements is necessarly an average over different meteorological conditions, and therefore it is a sum of non-Gaussian terms.\cite{sigi}

\subsection{Astrolabes and Heliometers in Europe and Brazil}

Solar astrolabes are small-aperture (10 cm) instruments observing the Sun in drift-scan mode. In the Calern Observatory at 1260 m near Nice Since 1975 the Sun has been monitored on a daily basis with a modified Danjon astrolabe, upgraded to the CCD semi-automated DORaySol.\cite{morand}
F. Laclare made observations along four decades, and contributed to spread this method in other countries. In Brasil also these observations started in 1970s in S\~{a}o Paulo and Rio de Janeiro. 
A. H. Andrei and his team in Rio de Janeiro Observat\'orio Nacional gathered the project in 2011 with the new mirror-heliometer.\cite{helio}
This heliometer, still in the 10 cm class instruments, allows instantaneous measurements of the solar diameter, at various position angles, while the astrolabe in the drift- scan method, requires from 2 to 6 minutes for each measurement.
With the heliometer the influence of low frequency seeing on image motion is greatly reduced, and only ordinary blurring and image stretching are effective. This not means that the heliometer gives instant measurements potentially seeing-free. The low frequency component of image's stretching acts on the data by enlarging their dispersion.
An analysis conducted on data from Huariou Solar Observing Station\cite{seeing} at the 10cm Full-Disc Vector Magnetograph \footnote{{\rm http://sun.bao.ac.cn/smct/r102\_e.html}} shows the role of the image's stretching over series of instant measurements of the radius lasting 15 minutes.

The statistical study of such disturbance can give new insights on the correct treatment of such data. 

\subsection{Heliolatitude and Solar Oblateness}
 
The heliometer is used to measure also the oblateness of the Sun, by changing the position angle of the Sun with a rotation of the telescope's axis.
The astrolabes, as all other methods based upon drift-scan transits, measure the solar diameter along different heliolatitudes, in dependance of the observing site and of the season. Since the oblateness effect is on the order of 8 mas\cite{RHESSI} this has to be taken into consideration when diameters measured at various heliolatitudes are averaged together.
The studies on the solar oblateness allow to propose models to interpret it: the combination of a nearly uniform rotating core with a prolate solar tachocline and an oblate surface.\cite{Rozelot} The measurements of the quadrupole term from the ground are possible, but of high difficulty and can be obtained only during excellent weather conditions. The hexadecapole term should be only obtained from space.\cite{Rozelot2} 
An astrometric satellite like Picard is expected to unambiguously determine the inertia moments of the Sun through the $J_n$ terms. 

\section{The Cycle 24}

In the prolonged minimum between cycles 23 and 24, we experienced 746 days with (0,0) counts at the Specola Solare Ticinese in Locarno from Jan 27, 2004 to Jan 15, 2011. 
These days are the first and the last spotless days of this minimum.\footnote{{\rm http://specola.ch/drawings/}  thanks to Sergio Cortesi and Marco Cagnotti for providing these data}
The Specola Solare is the observatory in Switzerland where the tradition of Wolf, Wolfer and Waldmeier on the observation of solar activity has been continued. The count of the number of spots and group (N, G) has been made rigorously in the same way since Wolf started, only with controlled modifications.
Now in the last years other Observatories, networked by the SIDC\footnote{{\rm http://www.sidc.be/}}, started to observe "more" spots.
Probably there is an influence of the space images, consulted by the observers before making the draw. This fact can increase the evaluation of the solar activity, with respect to a similar situation occurred in the past centuries, where all small spots or pores were not recorded because of the lack of space probes.
This debate is still ongoing between the Observatories of the SIDC, and it will give an important issue on the nature of this minimum, which can be considered indeed as the deepest since 1810-14.\footnote{{\rm http://www.metoffice.gov.uk/news/releases/archive/2012/solar-output-research}},\footnote{{\rm http://science.nasa.gov/science-news/science-at-nasa/2006/10may\_longrange/}},\footnote{D. Rose, http://www.dailymail.co.uk/sciencetech/article-2093264/Forget-global-warming--Cycle-25-need-worry-NASA-scientists-right-Thames-freezing-again.html (2012).} 

\begin{figure}
\centerline{\includegraphics[width=1\textwidth,clip=]{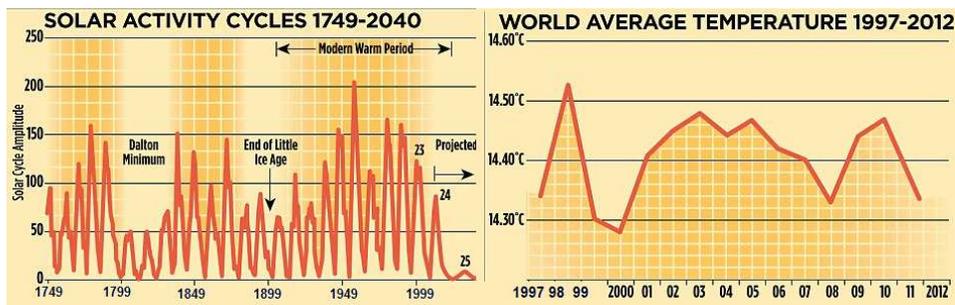}}
\caption{The prediction for solar cycle 25 and the World temperature in the last 15 years.$^{\rm{g}}$}
\label{Fig. 1}
\end{figure}

\section{The Sun and the Climate}

The solar influence on Earth's climate seems evident when we consider the Little Ice Age occurred during the Maunder Minimum.
Similar situations, but on shorter timescale, occurred during deep minima, like the Dalton minimum (1810-14).
I. Usoskin\cite{Usoskin} showed that either the status of Grand Minimum either the one of Grand Maximum for the Sun occur statistically on $\sim 20 \% $ of its life.
After 300 years a new Grand Minimum like the Maunder's one is rather probable.
Predictions in this sense have been made by several authors,\cite{penn,jiang,jiang2}.

Either connected or not with solar diameter issues, the influence of the Sun on our climate is of primary importance.\cite{Kou}
The diameter has to be considered as one of the indicators of the solar activity. In particular it is related through the equation

$W=dLogL_{\odot}/dLogR_{\odot}$

Our recent works\cite{sigi,eclipses} showed as naked eye eclipses
can be perceived in different ways from different observers, even if they are located on the edges of the shadow's path.
On the edges the duration of an eclipse changes rapidly with the distance from the true limit of the umbral shadow, so this distance can be evaluated by the perceived duration of the totality.
Even if the eclipse last a few seconds, if the 
duration is evaluated from a group of observers 
in the same location, it can vary of one while second. The French team experienced that situation in Hao atoll during the 2010 solar eclipse. The comparison with ephemerides showed similar situations for other accounts of naked eye modern observations available on the web, as the eclipse of 2008 in Alert, Nunavut.\footnote{{\rm http://www.cbc.ca/news/technology/story/2008/08/01/total-eclipse.html}}

\section{Past History of the Solar Diameter}
To understand the past solar activity, by assuming a constant value of W, past values of $R_{\odot}$ are required. These values have to be obtained from ancient eclipse data and from planetary transits.
\subsection{Ancient Eclipses}
Data on ancient eclipses available are: the eclipse of Clavius in Rome, 1567; the eclipse of Halley in 1715;
the eclipse of Newcomb in 1869 and the eclipse of Brown over New York City in 1925. For Clavius we guess the position near Piazza del Collegio Romano; for Halley one observer's position has been revised in the last 30 years; S. Newcomb of the US Naval Observatory gathered informations from volunteer observers spread all over the Country and only two of them were located by chance near the opposite edges; finally E. W. Brown of Yale University made a similar volunteer campaign and could find two witness at the edges in NY and in Providence.
Nowadays IOTA\footnote{{\rm http://www.lunar-occultations.com/iota/iotandx.htm}} members, skilled amateur astronomers,  made several coordinated observations of eclipses, worldwide, doing video with astrometric standards. IOTA members contributed greatly to the improvement of the method of measuring the solar diameter with eclipses.\cite{atlas,Raponi}

The search of the inflexion point of the limb darkening function LDF even for eclipse data, shows how the assumption of a LDF behaving like a step function for visual observations is na\"{i}ve.
Our studies are going to set an errorbar $\ge 0.2$ arcsec over the solar diameter recovered from past eclipses.

\subsection{Past and Future Planetary Transits}

I. I. Shapiro in 1980\cite{shapiro} used records of transits of Mercury to recover the past history of the solar diameter.\cite{svesh}
Further studies seems to confirm the constancy of the diameter within the errorbars.
Measurements made with different instruments, under perfect observing conditions, as in the case of Gambart and Bessel in 1832 yield different transit times, and different diameter of Mercury and, consequently, a different diameter of the Sun.\cite{Gambart,Bessel}
 
Nowadys the chord\cite{sigi,UAI} draft on the solar limb by the planet's disk can be recovered by photos, in conditions not affected by the black drop phenomenon.\cite{golub} 
The time in which the chord is zero, when the black drop is maximum, can be extrapolated from photographs made at 1s interval (and correctly chronodated) around the intermediate stages of ingress and egress. An improvement of this method is to follow the center of the planet, which is not affected by the atmosphere.\cite{venus2012}
After corrections for atmospheric refraction the ephemerides can be used to recover the solar diameter by comparison with the observed ingress and egress times.
The opportunity given by the forthcoming transit of Venus of 5/6 june 2012 and the one of Mercury of May 9, 2016 has to be exploited to do chronodated images of the Sun during the ingress and the egress phases of the transits.
In 2004 there are no useful chronodated images of the transit to measure the solar diameter.
Also the studies on the Venus aureole\cite{tanga} can be done with UTC synchronized high-resolution photos, in order to be exploited for solar diameter measurements.

\section{The solar mesosphere: what is the solar limb?}
The "new definition" of solar diameter during eclipses is now like the "old definition" of solar diameter with full disk images, already in use during the Dicke's experiments at Princeton on the solar oblateness.\cite{Dicke,Hill}
But, once again, the eclipses are raising the problem on what is exactly the solar limb?
Because the inflexion point of the limb darkening function is a theoretical concept.
The Sun is a selfgravitating sphere of gas, and asking were is the border it is like to ask were is the border of the clouds when we fly over: there is a region where the Sun ends not a sharp edge.
Moreover at the limb the light is coming both from the continuum of the Thomson scattering from the photosphere, and from a forest of tiny emission lines, coming from slightly above the photosphere.
This has been evidenced in flash spectrum experiments since 1905.\cite{bazin}
The contribution of such region, called solar mesosphere in analogy with the Earth's atmosphere, is a blend of white light. It has been confused with the photosphere during video made with filters of different densities, and it could have been the responsible of the annular appearance of the eclipse of Clavius, which would have been hybrid, total in Rome, according to modern ephemerides.
With flash spectra the true continuum can be detected among the emission lines. The light curve of the true continuum produced by a Baily's bead (the last one in case of photometric, unidimensional, observations) can be deconvolved by using the Kaguya profile of the lunar valley producing the bead, to obtain the LDF and its inflexion point.

\section{Conclusions}

\begin{itemize}
\item{To study the past climate on the Earth, the past solar irradiance can be investigated by knowing the actual $W$ parameter and the past values of the solar diameter.}

\item{One of the goal of the French space mission Picard (now on duty) is to measure W, as well as of the SDS balloon borne mission (with flights in 1990s and 2009 and data of irradiation from ACRIM satellite).}

\item{The solar diameter is therefore an indicator of the solar activty in term of its irradiance.
Its monitoring helps us to better understand the Sun and the energy balance between gravitational and magnetic energy in the outer layers of our star.}

\item{Past measurements of the solar diameter, with ancient eclipses, are being rediscussed after having understood the role of the emission lines visible in flash spectra.}

\item{The use of the inflexion point of limb darkening function both in eclipse data and full disk observations is a conceptual unification of the methods.}

\item{Eclipses and planetary transits remain the most accurate method from Earth, but they are rare, while drift-scan transits or heliometric instant measurements are less accurate because of the influence of low frequency seeing acting both as image motion and stretching. The low frequency seeing influences for drift-scan and heliometric groundbased measurements.}

\end{itemize}

\end{document}